\documentstyle[preprint,aps]{revtex}
\begin{document}
\title{Total dielectric function approach to the electron Boltzmann
equation for scattering from a two-dimensional coupled mode system}
\author{B.A. Sanborn}
\address{Center for Solid State Electronics Research \\
Arizona State University, Tempe, Arizona \,\, 85287-6206}
\maketitle
\begin{abstract}
The nonequilibrium total dielectric function
lends itself
to a simple and general method for calculating the inelastic collision term in
the electron Boltzmann equation for scattering from a coupled mode system.
Useful applications include scattering from plasmon-polar phonon hybrid modes
in modulation doped semiconductor
structures.
This paper presents numerical methods for including inelastic
scattering at momentum-dependent hybrid phonon frequencies in the
low-field Boltzmann equation for two-dimensional electrons coupled to
bulk phonons.
Results for
mobility in GaAs
show that the influence of
mode coupling and dynamical screening on
electron scattering from polar optical phonons is stronger
for two dimensional electrons than was previously found for the three
dimensional case.
\end{abstract}

\section{Introduction}
The importance of hybridization of collective modes has been pointed out
in the context of several semiconductor systems. Examples include
hybrid optical modes
in thin semiconductor layers\cite{RIDLEY},
coupled intraband-interband excitations of
quasi-one-dimensional electron systems\cite{LI}, and, more commonly,
plasmon--phonon coupled modes in doped polar semiconductors
\cite{COUP,SAN95a,SAN95b}.
To the extent that electron scattering depends on the energies and interaction
strengths of the system modes, a reliable theory of electron transport
depends on treating mode-coupling effects accurately.

In the Born approximation, the interaction between a conduction electron and
a coupled electron-phonon system can be represented as an effective
electron-electron interaction screened by the nonequilibrium total
dynamic dielectric function
$\epsilon_{T}(q,\omega)$, which includes contributions from both electrons
and phonons.
Recent work\cite{SAN95a,SAN95b} shows that
$\epsilon_{T}(q,\omega)$ provides a systematic way to determine
the collision term
in the electron Boltzmann equation for
scattering against dynamically screened coupled electron-phonon modes.
The result is the
sum of an electron-electron collision term and an electron-LO phonon
collision term that includes plasmon-phonon mode coupling.
Both interactions are dynamically screened by only the electronic part of the
total dielectric function for the electron-phonon system.
In the random-phase approximation (RPA), the electron-phonon part
contains a phonon self-energy that arises from the polarization of the
electron gas.\cite{MAHAN} The self-energy correction modifies the
longitudinal-optical (LO)-phonon dispersion in doped polar semiconductors,
producing hybrid normal modes with phonon strength in each.
Numerical results\cite{SAN95b} for bulk n-type GaAs show that
mode-coupling and dynamical screening should significantly influence
electron mobilities in modulation doped structures.

The present paper
describes numerical methods for exactly solving the
low-field Boltzmann equation for two-dimensional electrons
coupled to bulk LO phonons,
including dynamical screening and mode coupling
for arbitrary electron degeneracy  and spherical
energy surfaces. In particular, the method for including phonon
dispersion in the collision integrals is discussed.
The phonon distribution is approximated
 by its equilibrium form, and the plasmon-pole
approximation\cite{PP} is used in the LO phonon self-energy
to determine the hybrid mode frequencies.
Results for GaAs  show that
mode coupling and dynamical screening are more important for two
dimensional electrons than for the three dimensional case.

\section{Inelastic scattering from hybrid phonons with dispersion}

The collision integrals in the electron Boltzmann equation
for doped polar semiconductors contain the
momentum-dependent frequencies of the hybrid LO phonon
modes\cite{SAN95a}.
Using the plasmon-pole approximation in the phonon self-energy,
the differential scattering rate
due to transfer of momentum
${\bf q}={\bf k}-{\bf p}$ to the hybrid phonons
is\cite{SAN95a,SAN95b}
\begin{eqnarray}
W^{LO}({\bf k,p})=\frac{-2M_{q}^{2}}{\Omega\hbar|\epsilon(q,\omega_{k,p})|^{2}}
[N(\omega_{k,p})+1]{\rm Im}[D^{+}(q,\omega_{k,p})+D^{-}(q,\omega_{k,p})]
\label{WLO} \\
{\rm Im}[D^{\pm}(q,\omega)]=\mp\frac{\pi \omega_{TO}(\omega^{2}-\tilde{\omega}
_{p}^{2})}{\hbar \omega_{\pm}(\omega_{+}^{2}-\omega_{-}^{2})}[\delta(\omega+
\omega_{\pm})-\delta(\omega-\omega_{\pm})] \,,\label{IMDPM}
\end{eqnarray}
where $M_{q}^{2}=v_{q}(\epsilon_{\infty}^{-1}-\epsilon_{0}^{-1})\hbar\omega_
{LO}^{2}/2\omega_{TO}$ and $\hbar\omega_{kp}=E_{k}-E_{p}$.
Here, $\omega_{LO}$ and $\omega_{TO}$ are the LO and transverse optical
phonon frequencies, while $\epsilon_{\infty}$ and
$\epsilon_{0}$ are the high-frequency and static dielectric constants,
respectively.
In two dimensions, $v_{q}=2\pi e^{2}/q\epsilon_{\infty}$.
The frequencies $\tilde{\omega}_{p}$, $\omega_{+}$, and $\omega_{-}$ are
given by
\begin{eqnarray}
\tilde{\omega}_{p}^{2}=\omega_{p}^{2}[1-\epsilon^{-1}(q,0)]^{-1}\\
\omega^{2}_{\pm}=\frac{1}{2}\left\{\omega^{2}_{LO}+\tilde{\omega}^{2}_{p}\pm
\left[(\omega_{LO}^{2}-\tilde{\omega}^{2}_{p})^{2}+4\omega_{p}^{2}(\omega_{LO}^
{2}-\omega_{TO}^{2})\right]^{1/2}\right\}\,,\label{OMEGAPM}
\end{eqnarray}
where $\omega_{p}=(2\pi ne^{2}q/m^{*}\epsilon_{\infty})^{1/2}$ is the plasmon
frequency of the two-dimensional electron gas with concentration $n$
and effective mass $m^{*}$.
The weight factors $(\omega_{\pm}^{2}-\tilde{\omega}_{p}^{2})(\omega_{+}^{2}-
\omega_{-}^{2})^{-1}$ in Im$[D^{\pm}]$ give the phonon strength in each of
the hybrid $\omega_{\pm}$ modes, so that the differential scattering rate
$W^{LO}$ is the rate for scattering from only the phonon component of the
hybrid modes.
The screening function in (\ref{WLO}) is the temperature-dependent
RPA dielectric function, $\epsilon(q,\omega)=1-v_{q}^{\infty}P(q,\omega)$,
for the two-dimensional electron gas.

The iterative procedure used here
for solving the low-field Boltzmann equation was described in reference
\cite{SAN95b}.
One assumes a linear form for the nonequilibrium electron distribution,
$f({\bf k})=f^{0}_{k}+x_{k}g_{k}$,
where $f^{0}_{k}$ is the equilibrium (Fermi-Dirac) distribution,
$x_{k}$ is the cosine of the angle between the electric field
${\bf F}$ and ${\bf k}$,
and $g_{k}$
is an unknown function that is linear in $F$.
Neglecting
all scattering mechanisms except electron-LO-phonon gives
\begin{eqnarray}
g_{k}=\left[\nu_{LO}[g]-\frac{eF}{\hbar}\frac{\partial f^{0}_{k}}
{\partial k}\right]\left[\tau^{-1}_{LO}(k) \right]^{-1} \,,\\
\tau^{-1}_{LO}(k)=\frac{1}{\Omega}\sum_{{\bf p}}\bigl\{W^{LO}({\bf k,p})
[1-f^{0}_{p}]+
W^{LO}({\bf p,k})f^{0}_{p}\bigr\} \,,\\
\nu_{LO}[g]=\frac{1}{\Omega}\sum_{{\bf p}}
g_{p}x_{kp}\bigl\{W^{LO}({\bf p,k})[1-f^{0}_{k}]+
W^{LO}({\bf k,p})f^{0}_{k}\bigr\} \,,
\end{eqnarray}
and $x_{kp}$ is the cosine of the angle between ${\bf k}$ and ${\bf p}$.

The form of the phonon Green's function in equation (2) implies that
the energy conservation relation between initial and final
electron states,
$E_{k}-E_{{\bf k}-{\bf q}}=\pm\hbar \omega (q)$, is dependent on
the momentum transfer $q$ through the phonon dispersion.
In three dimensions\cite{SAN95b}, it is convenient to
choose $q$ as an integration
variable when calculating $\nu_{LO}$ and $\tau^{-1}_{LO}$. In this
case, one must determine the integration limits by finding the points
where the phonon dispersion curve $\omega(q)$ intersects the
absorption region $q^{2}+2kq \leq 2m^{*}\omega(q)/\hbar \leq
q^{2}-2kq$ and the emission region $2m^{*}\omega(q) \leq -q^{2}
+2kq$. For the case of two-dimensional electrons, it appears that
$q$ is not a preferred choice for integration variable, since the
integrand then diverges at the integration limits. Instead one
can choose $\theta_{kp}$, the angle between ${\bf k}$ and ${\bf p}$.
In this case, one must determine the implicit relation between
$q$ and $\theta_{kp}$ at each integration step by finding the
(nonzero) root of the function
$f(q)=q^{2}-k^{2}-p_{\pm}^{2}+2kp_{\pm} \cos \theta_{kp}$
with $p_{\pm}=(k^{2}\pm 2m^{*}\omega_{\pm}/\hbar)^{1/2}$.

Scattering from the LO phonon part of the hybrid modes gives
the inverse lifetime $\tau^{-1}_{LO}(k)$ and functional $\nu _{LO}[g]$,
\begin{eqnarray}
\tau^{-1}_{LO}(k)=
\sum_{\lambda=\pm}
\lambda&\Biggl\{\Biggr.&\int_{0}^{2\pi}
\,d\theta_{kp} I(q,\omega_{\lambda})
\Bigl[N(\omega_{\lambda})+1-f^{0}_{k_{\lambda}^{-}}\Bigr]
\Theta(E_{k}-\hbar\omega_
{\lambda}) \nonumber \\
& &+\int_{0}^{2\pi}
\,d\theta_{kp} I(q,\omega_{\lambda})
\Bigl[N(\omega_{\lambda})+f^{0}_{k_{\lambda}^{+}}\Bigr]\Biggr\} \,,
\label{TAULO} \\
\nu_{LO}[g]=
\sum_{\lambda=\pm}
\lambda&\Biggl\{\Biggr.&\int_{0}^{2\pi}
\,d\theta_{kp} I(q,\omega_{\lambda})
g(k_{\lambda}^{-})
\cos (\theta_{kp})
\Bigl[N(\omega_{\lambda})+f^{0}_{k}\Bigr]
\Theta(E_{k}-\hbar\omega_
{\lambda}) \nonumber \\
& &+\int_{0}^{2\pi}
\,d\theta_{kp} I(q,\omega_{\lambda})
g(k_{\lambda}^{+})\cos (\theta_{kp})
\Bigl[N(\omega_{\lambda})+1-f^{0}_{k}\Bigr]\Biggr\}\,,\label{NULO}
\end{eqnarray}
where
$k_{\lambda}^{\pm}=(k^{2}\pm2m^{*}\omega_{\lambda}/\hbar)^{1/2}$ and
\begin{displaymath}
I(q,\omega_{\lambda})=\frac{e^{2}\omega_{LO}^{2}m^{*}}{2q\omega_{\lambda}
\hbar^{2}}
(\frac{1}{\epsilon_{\infty}}-\frac{1}{\epsilon_{0}})\mid\epsilon
(q,\omega_{\lambda})\mid^{-2}
\frac{(\omega^{2}_{\lambda}-\tilde{\omega}_{p}^{2})}{(\omega_{+}^{2}-
\omega_{-}^{2})} \,.
\end{displaymath}

\section{Results and conclusions}
Figures 1 and 2 present electron mobility as a function of
concentration in GaAs at 300 K and 77 K, showing
the effects of dynamic screening and mode coupling
on the electron-LO phonon interaction.
The mobilities were calculated by averaging
the electron velocity
over the nonequilibrium distribution $f({\bf k})$,
found by solving the iterative equation (5).
Without plasmon-phonon coupling, dynamic RPA screening gives
mobilities that are lower than static RPA or
Thomas-Fermi results, but higher than the unscreened
case.
The difference
between  static and dynamic screening has a strong momentum
dependence.
The inverse dynamic RPA dielectric function evaluated at $\omega=
\omega_{LO}$ has a strong peak
($\epsilon ^{-1} > 1$) at small $q$ where the
plasmon dispersion crosses $\omega_{LO}$, but is smaller
than the inverse static dielectric function
at larger $q$ values.
The small $q$ values, where antiscreening and enhanced {\em forward}
scattering
occur, are accessible only for high energy electrons, since
the minimum $q$ value allowed by energy conservation is
$q_{min}=|k-(k^{2}+2m^{*}\omega_{LO}/\hbar)^{1/2}|$. Previous
authors\cite{YANG} have noted similar effects in calculations of
dynamical screening of the electron-LO phonon interaction.

When plasmon-phonon coupling is included in the
plasmon-pole model, the phonon spectral function $-\pi ^{-1}
{\rm Im}[D(q,\omega)]$ has two peaks
with frequencies $\omega_{+}$ and $\omega_{-}$ given by
(\ref{OMEGAPM}).
Mobilities are expected to be lower when
mode coupling is included, because of
increased low-energy electron-phonon scattering due to the
$\omega_{-}$ mode.
Figures 1 and 2 show that this is in fact the case.
As in three dimensions\cite{SAN95b}, mode coupling
works to reduce electron mobility especially
for low densities at 77K
where the thermal occupation of the low-energy hybrid
mode is exponentially larger than the high-energy hybrid
mode or uncoupled phonon mode.
The effect is more pronounced in
two dimensions, consistent with previous conclusions
concerning the effect of dimensionality on polaronic
damping.\cite{MASON}

\newpage
\begin{figure}
\caption{Effects of screening and mode coupling on LO phonon limited
mobility of two-dimensional electrons
in GaAs at 300K. The dynamically screened
coupled mode mobility calulation (solid curve) is compared to
mobilities determined by scattering from uncoupled LO phonons screened
in the Thomas-Fermi approximation(long-dashed curve), static RPA
(short-dashed curve), dynamic RPA
(dotted curve), and unscreened (dot-dashed curve). All other scattering
mechanisms are neglected.}
\label{fig1}
\end{figure}

\begin{figure}
\caption{Effects of screening and mode coupling on LO phonon limited
mobility of two-dimensional electrons in GaAs at 77K.
Mobility curves are represented as in Figure 1.}
\label{fig2}
\end{figure}

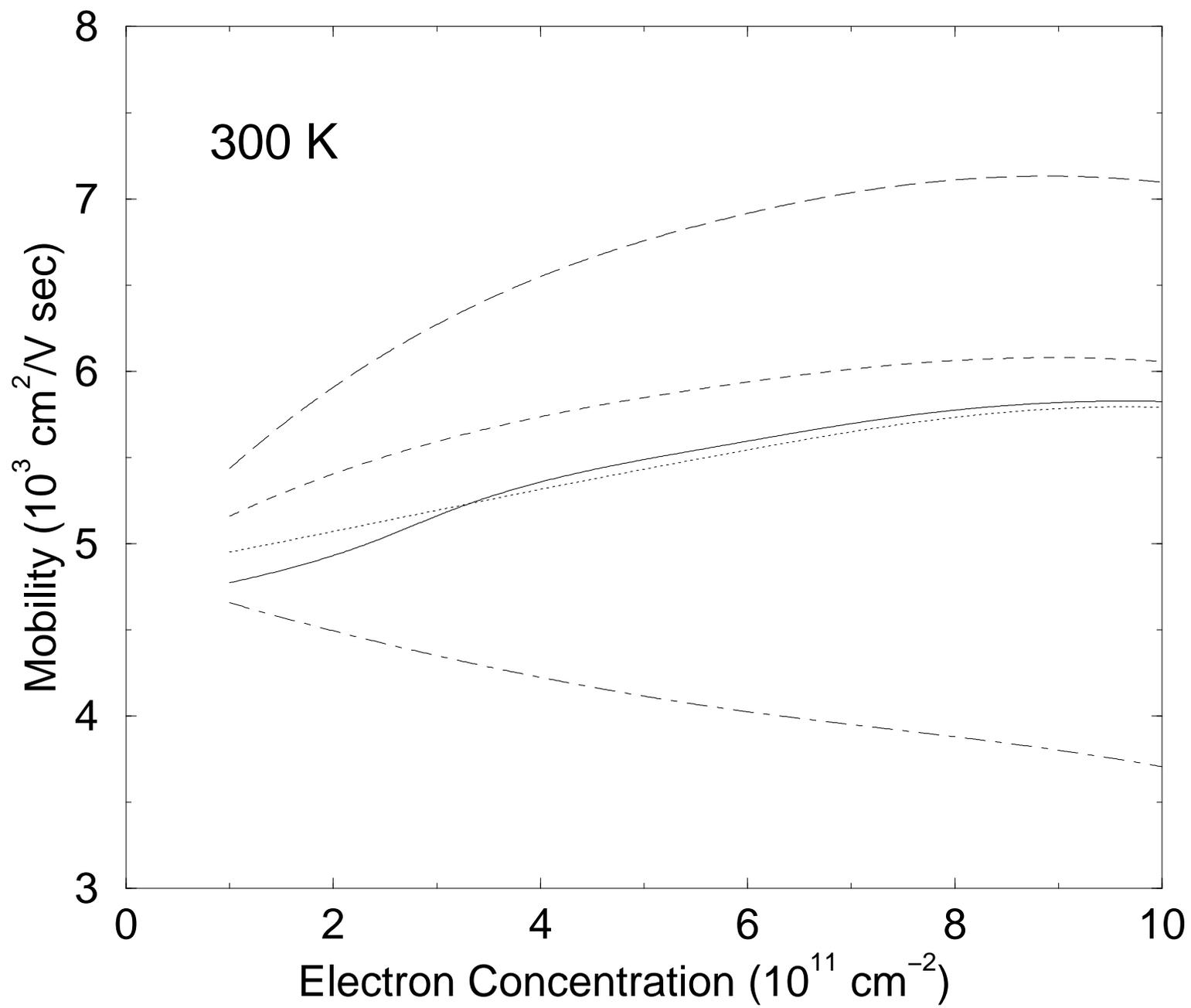
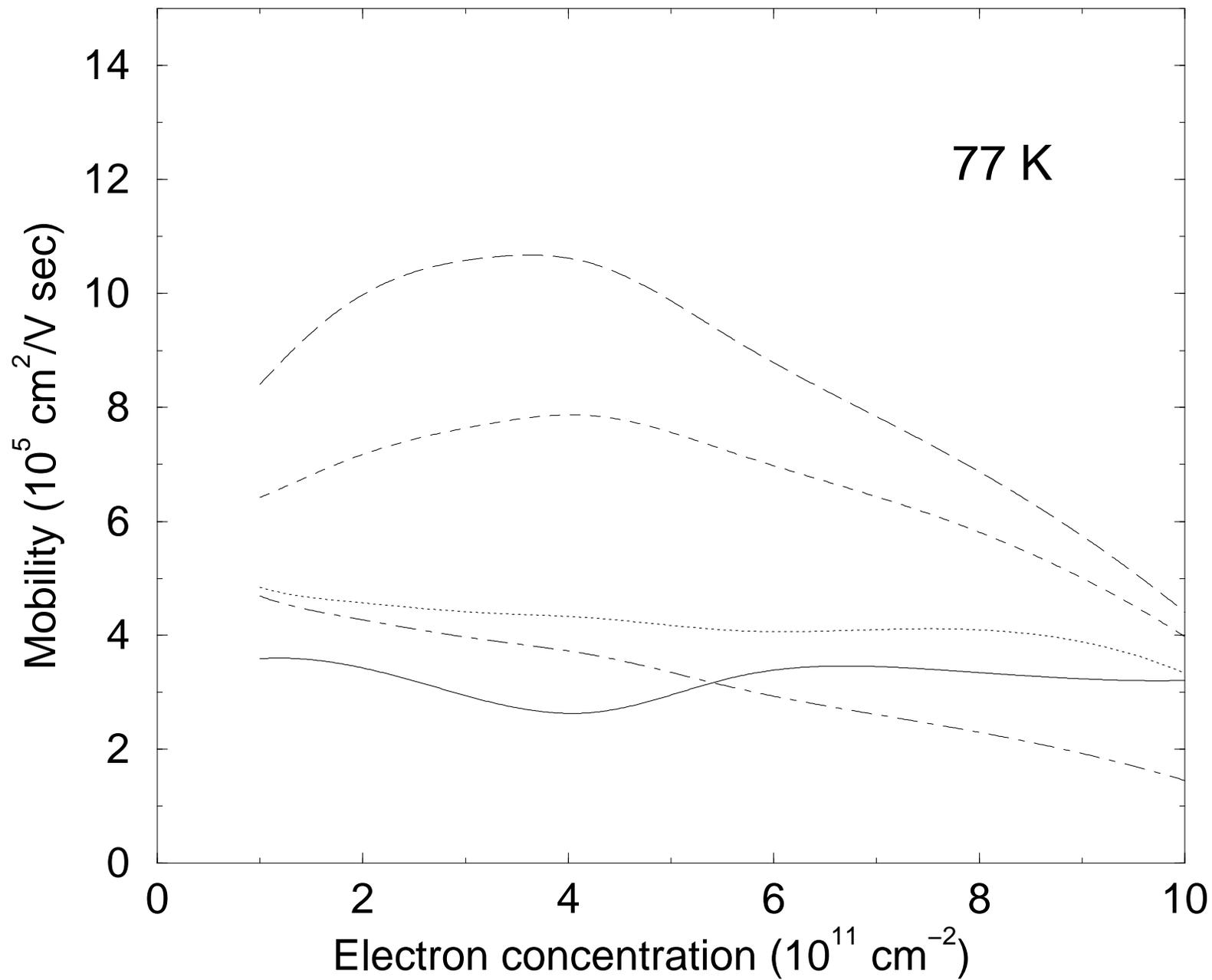
\end{document}